# Artificial Intelligence-Enabled Intelligent 6G Networks

Helin Yang, *Student Member, IEEE*, Arokiaswami Alphones, *Senior Member, IEEE*, Zehui Xiong, Dusit Niyato, *Fellow, IEEE*, Jun Zhao, *Member, IEEE*, and Kaishun Wu, *Senior Member, IEEE*

## Abstract

With the rapid development of smart terminals and infrastructures, as well as diversified applications (e.g., virtual and augmented reality, remote surgery and holographic projection) with colorful requirements, current networks (e.g., 4G and upcoming 5G networks) may not be able to completely meet quickly rising traffic demands. Accordingly, efforts from both industry and academia have already been put to the research on 6G networks. Recently, artificial intelligence (AI) has been utilized as a new paradigm for the design and optimization of 6G networks with a high level of intelligence. Therefore, this article proposes an AI-enabled intelligent architecture for 6G networks to realize knowledge discovery, smart resource management, automatic network adjustment and intelligent service provisioning, where the architecture is divided into four layers: intelligent sensing layer, data mining and analytics layer, intelligent control layer and smart application layer. We then review and discuss the applications of AI techniques for 6G networks and elaborate how to employ the AI techniques to efficiently and effectively optimize the network performance, including AI-empowered mobile edge computing, intelligent mobility and handover management, and smart spectrum management. Moreover, we highlight important future research directions and potential solutions for AI-enabled intelligent 6G networks, including computation efficiency, algorithms robustness, hardware development and energy management.

H. Yang and A. Alphones are with the School of Electrical and Electronic Engineering, Nanyang Technological University, Singapore 639798 (e-mail: hyang013@e.ntu.edu.sg, ealphones@ntu.edu.sg).

Z. Xiong, D. Niyato, and J. Zhao are with the School of Computer Science and Engineering, Nanyang Technological University, Singapore 639798 (e-mail: zxiong002@e.ntu.edu.sg, dniyato@ntu.edu.sg, junzhao@ntu.edu.sg).

K. Wu is with the College of Computer Science and Software Engineering, Shenzhen University, Shenzhen 518060, China (e-mail: wu@szu.edu.cn).

# I. INTRODUCTION

Wireless networks have evolved from the first generation (1G) networks to the upcoming/recent fifth generation (5G) networks with various attentions like data rate, end-to-end latency, reliability, energy efficiency, coverage, and spectrum utilization. According to the International Telecommunication Union (ITU), 5G networks have three main types of usage scenarios: enhanced mobile broadband (eMBB), ultra-reliable and low latency communication (URLLC) and massive machine-type communications (mMTC) to account for supporting diverse services [1], [2]. In that regard, technologies including millimeter-wave (mmWave), massive multiple-input multiple-output (MIMO), device-to-device (D2D) technologies, and so on, are employed to provide users with better quality of service (QoS) and quality of experience (QoE), as well as improve the network performance [1], [2].

While 5G networks are being deployed, people from both the industry and academia have already paid attention to the research on 6G networks [3]-[7], where 6G networks are expected to effectively support high-quality services, new emerging applications (e.g., virtual and augmented reality, remote surgery, and holographic projection) and unlimited connectivity for the massive number of smart terminals. For instance, the references [3] and [4] discussed the roadmap toward 6G networks along with requirements, enabling techniques and architectures.

Different from previous generation networks, 6G networks will be required to revolutionize themselves by realizing intelligence to meet more stringent requirements and demands for the intelligent information society of 2030, which include [3]-[6]: ultrahigh data rates, a peak data rate of at least 1 Tb/s and a user-experienced data rate of 1 Gb/s; ultralow latency, less than 1 ms end-to-end delay, even 10–100 μs; ultrahigh reliability, about $1-10^{-9}$; high energy efficiency (EE), on the order of 1 pJ/b; very high mobility, up to 1000 km/h; massive connection, up to $10^7$ devices/km$^2$ and traffic capacity of up to 1 Gbs/m$^2$; large frequency bands (e.g., 1THz-3THz); connected intelligence with AI capability.

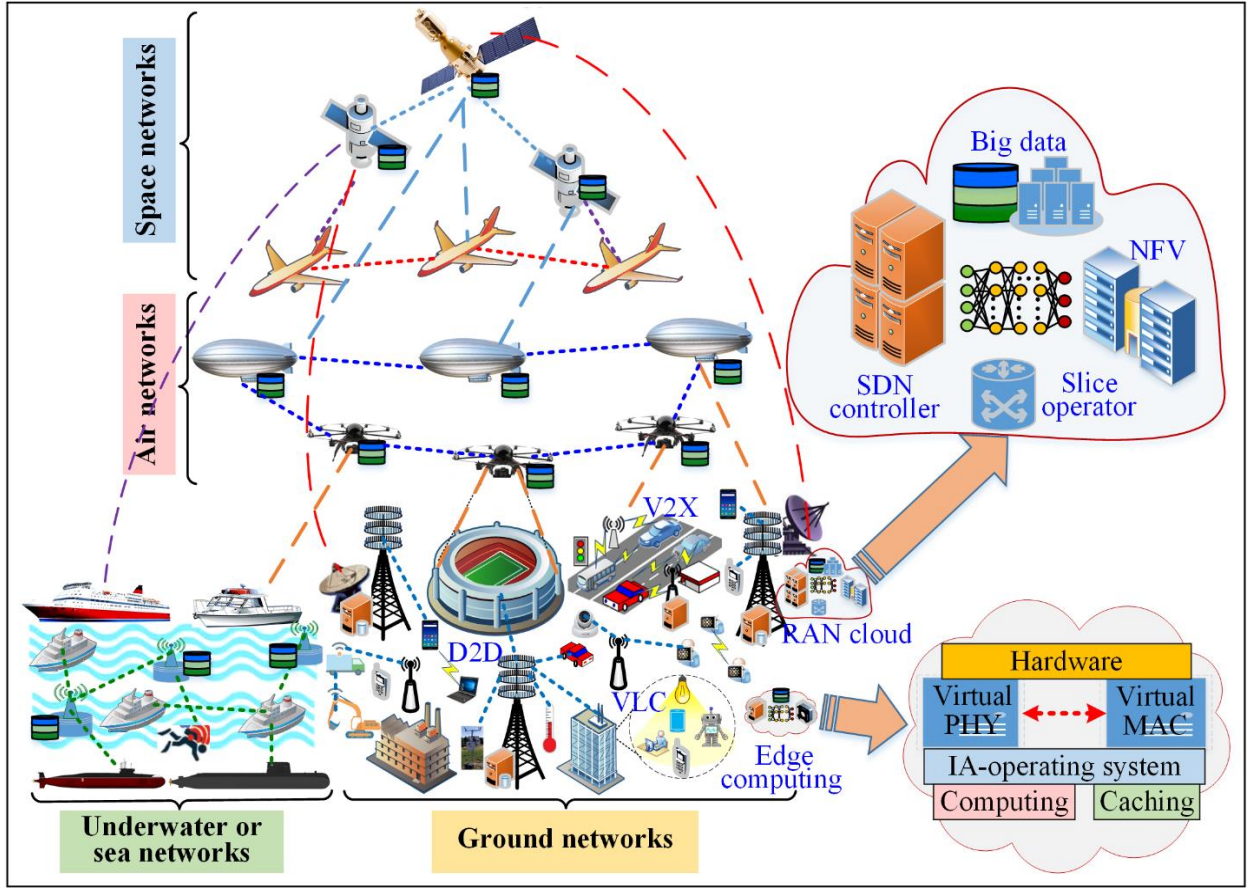

Figure 1. The typical architecture of 6G network (ISAGUN). V2X: vehicle to everything; VLC: visible light communication; RAN: radio access networks; SDN: software-defined networking; NFV: network function virtualization; PHY: physical layer; MAC: medium access control.

Note: The objective of ISAGUN is to extremely broad coverage and seamless connectivity for space, airborne, ground, and underwater areas, such as flight in the sky, ship at sea, monitoring at remote areas or vehicles on land. As a result, human activity will dramatically expand from the ground to air, space, and deep sea. At the same time, centralized and edge computing are deployed at RAN with SND and NFV to provide powerful computational processing and massive data acquisition for ISAGUN.

Furthermore, in order to support near-instant and seamless super-connectivity, an integrated space–air–ground–underwater network (ISAGUN) will be the core potential architecture of 6G networks [6], [8], as shown in Fig. 1, where ISAGUN mainly consists of the following four tiers:

- Space-network tier deploys low Earth orbit, medium Earth orbit, and geostationary Earth orbit satellites [8] to provide orbit or space services for unserved areas which are not covered by ground networks.
- Air-network tier employs various aerial platforms including unmanned aerial vehicles

- (UAVs), airships, and balloons associated with flying base stations (BSs) to support flexible and reliable wireless connectivity for remote areas or urgent events.
- Ground-network tier is the main solution for supporting diverse services for a massive number of devices. In order to satisfy various services, this layer mainly exploits low-frequency, microwave, mmWave, visible light and Terahertz (THz) bands for 6G networks.
- Underwater-network tier aims to provide underwater communication connectivity, observation and monitoring services for broad-sea and deep-sea activities.

According to the former evolution rules of networks, initial 6G networks will be mainly supported by the existing 5G infrastructures, such as the architectures of SDN, NFV and network slicing (NS). However, compared with 5G networks, 6G networks require to support the above mentioned stringent requirements (e.g., ultrahigh data rates, ultralow latency, ultrahigh reliability, and seamless connectivity). At the same time, the development of 6G networks (ISATUN) has large dimension, high complexity, dynamicity and heterogeneity characteristics. All the above mentioned issues call for a new architecture that is flexible, adaptive, agile and intelligent. Artificial intelligence (AI) [7], [9], with strong learning ability, powerful reasoning ability and intelligent recognition ability, allows the architecture of 6G networks to learn and adapt itself to support diverse services accordingly without human intervention. In [5], AI-enabled techniques were applied to achieve network intelligentization, closed-loop optimization and intelligent wireless communication for 6G networks. Kato *et al.* [8] used deep learning to optimize the performance of Space-Air-Ground Integrated networks, and showed how to employ deep learning to select most suitable paths for satellite networks. Furthermore, deep reinforcement learning (DRL) was adopted to preserve reliable wireless connectivity for UAVs-enabled networks by learning the environment dynamics [10], and simulation demonstrated that DRL significantly outperforms conventional methods. Hence, it is promising to adopt AI to 6G networks to optimize the network architecture and improve the network performance.

Although the references [5], [7] and [8] applied AI to enable intelligent wireless networks, they did not investigate how to systematically sense data from environments, analyze collected data, and then apply the discovered knowledge to optimize network performance for 6G. Hence, in this article, an AI-enabled intelligent architecture for 6G networks is presented to realize smart resource management, automatic network adjustment and intelligent service provisioning with a high level of intelligence, where the architecture consists of four layers: sensing layer, data mining and analytics layer, control layer and application layer. The proposed architecture is capable of intelligently extracting valuable information from massive data, learning and supporting different functions for self-configuration, self-optimization, and self-healing in 6G networks, in order to tackle optimized physical layer design, complicated decision making, network management and resource optimization tasks. Based on AI-enabled intelligent 6G networks, we introduce the applications of AI techniques in the context of AI-empowered mobile edge computing, intelligent mobility and handover management, and smart spectrum management. After that, we discuss important future research directions for AI-enabled 6G intelligent networks. Finally, we conclude this article.

## II. AI-ENABLED INTELLIGENT 6G NETWORKS

The development of 6G networks will be large-scale, multi-layered, high complex, dynamic, and heterogeneous. In addition, 6G networks need to support seamless connectivity and guarantee diverse QoS requirements of the huge number of devices, as well as process large amount of data generated from physical environments. AI techniques with powerful analysis ability, learning ability, optimizing ability and intelligent recognition ability, which can be employed into 6G networks to intelligently carry out performance optimization, knowledge discovery, sophisticated learning, structure organization and complicated decision making. With the help of AI, we present an AI-enabled intelligent architecture for 6G networks which is mainly divided into four layers: intelligent sensing layer, data mining and analytics layer, intelligent control layer and smart application layer, as shown in Fig. 2.

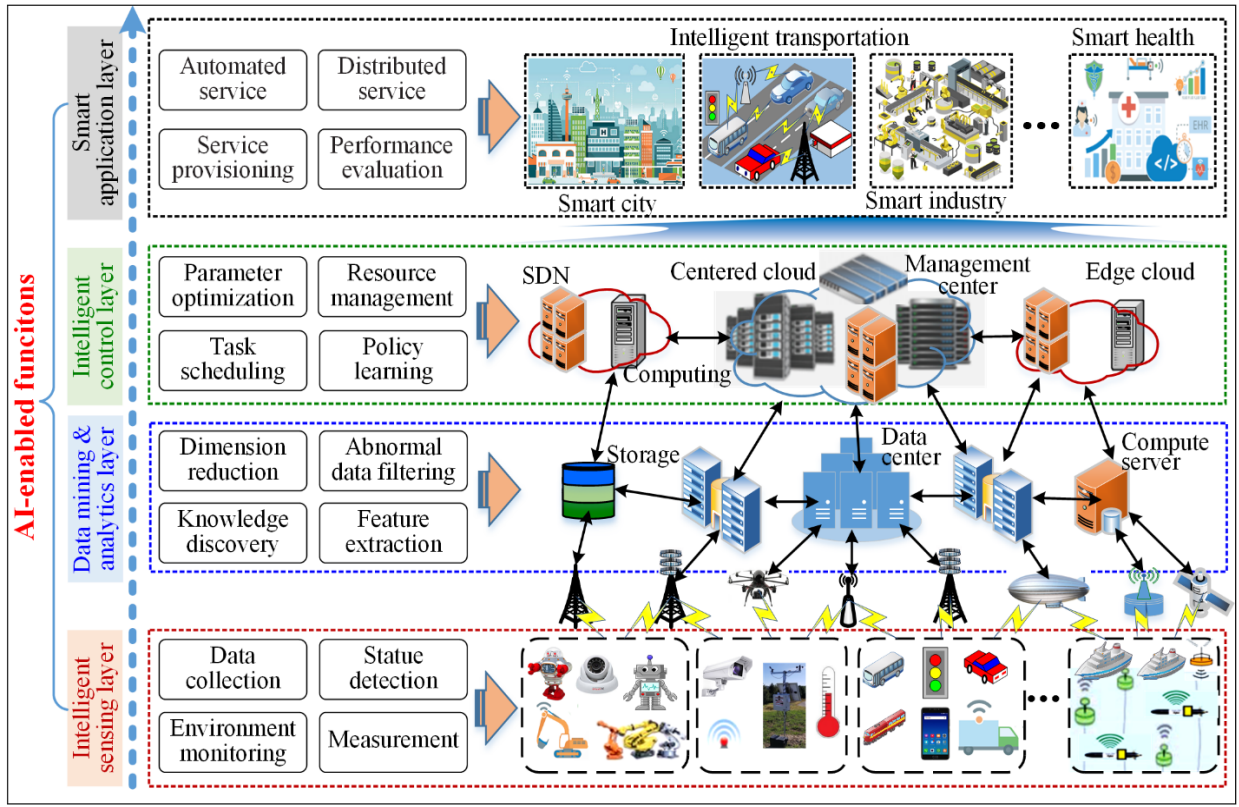

Figure 2. The architecture of AI-enabled intelligent 6G networks.

Here, we introduce some common AI techniques as follows first. AI techniques subsume multi-disciplinary techniques including machine learning (supervised learning, unsupervised learning and reinforcement learning), deep learning, optimization theory, game theory, and meta-heuristics [9]. Among them, machine learning and deep learning are most popular AI subfields which are widely adopted in wireless networks.

**Supervised Learning:** Supervised learning uses a set of exclusive labeled data to build the learning model (also called training), which is broadly divided into classification and regression subfields. Classification analysis aims to assign a categorical label to every input sample, which mainly includes decision trees (DT), support vector machine (SVM) and K-nearest neighbors (KNN). Regression analysis contains support vector regression (SVR) and Gaussian process regression (DPR) algorithms, and it estimates or predicts continuous values based on the input statistical features.

**Unsupervised Learning:** The task of unsupervised learning is to discover hidden patterns as well as extract the useful features from unlabeled data, and it is generally divided into clustering and dimension reduction. Clustering seeks to group a set of samples into different clusters according to their similarities, and it mainly includes K-means clustering and hierarchical clustering algorithms. Dimension reduction transforms a high-dimensional data space into a low-dimensional space without losing much useful information. Principal component analysis (PCA) and isometric mapping (ISOMAP) are two classic dimension reduction algorithms.

**Reinforcement Learning (RL):** In RL, each agent learns to map situations to actions and makes suitable decisions on what the actions to take through interacting with the environment, so as to maximize a long term reward. Classic RL algorithms include Markov decision process (MDP), Q-learning, policy learning, actor critic (AC), DRL and multi-armed bandit (MRB).

**Deep Learning:** Deep learning is an AI function that realizes the working of the human brain in understanding the data representations and creating patterns based on artificial neural networks. It consists of multiple layers of neurons, and the learning model can be supervised, semi-supervised and unsupervised. Classic deep learning algorithms include deep neural network (DNN), convolutional neural network (CNN), recurrent neural network (RNN) and long short-term memory (LSTM).

*A. Intelligent Sensing Layer*

Generally, sensing and detection are the most primitive tasks in 6G networks, where 6G networks tend to intelligently sense and detect the data from physical environments via enormous devices (e.g., cameras, sensors, vehicles, drones and smartphones) or crowds of human beings. AI-enabled sensing and detecting are capable of intelligently collecting the large amounts of dynamic, diverse and scalable data by directly interfacing the physical environment, mainly including radio-frequency utilization identification, environment monitoring, spectrum sensing, intrusion detection, interference detection, and so on.

It is worth noticing that high accurate sensing, real-time sensing and robust sensing are of great interest, since 6G networks need to support ultrahigh reliability and ultralow latency communication services. In addition, dynamic 6G networks lead to spectrum characteristic uncertainty, which entails great difficulty in robust and accurate sensing. AI techniques can realize accurate, real-time and robust spectrum sensing, where fuzzy SVM and nonparallel hyperplane SVM are robust to the environment uncertainties, CNN based cooperative sensing can improve sensing accuracy with low complexity, the combination of $K$-means clustering and SVM is capable of achieving real-time sensing by training low-dimensional input samples, and Bayesian learning can address large-scale heterogeneous sensing problems by tackling heterogeneous data fusion.

For example, in 6G networks, spectrum sensing is an important technique to improve the spectrum usage efficiency and address spectrum scarcity problems. However, spectrum sensing in large-scale 6G networks is very challenging since a massive number of devices aim to sense spectrum simultaneously, and the massive spectrum usage detections lead to high-dimensional search problems. In this case, AI technologies can be applied to identify the spectrum characteristics, and intelligently establish suitable training models to sense spectrum working status. In detail, AI-enabled learning models (e.g., SVM and DNN) detect the spectrum working status by categorizing each feature vector (spectrum) into either of the two classes, namely, the "spectrum idle class" and "spectrum buy class", and adaptively update the leaning models based on dynamic environments.

*B. Data Mining and Analytics Layer*

This layer is a core task that aims to process and analyze the massive amounts of raw data generated from the huge number of devices in 6G networks, and achieve semantic derivation and knowledge discovery. The massive collected data from physical environments may be heterogeneous, nonlinear, and high dimensional, so data mining and analytics can be applied in 6G networks to address the challenges of processing the massive amount of data, as well as to analyze the collected data towards knowledge discovery.

On the one hand, it is costly to transmit or store the massive raw data in dense networks. Hence, it is necessary to reduce data dimension of the raw data, filter abnormal data, and finally achieve a more reasonable dataset. AI-based data mining, such as PCA and ISOMAP are two common AI algorithms which can help 6G networks to transform higher-dimensional data into a lower-dimensional subspace [7], which dramatically decreases the computing time, storage space and model complexity. For example, the massive collected channel information, traffic flows, images and videos from sensors, devices or social media are high-dimensional data, which can be compressed into a small set of useful variables of the raw data by using PCA or ISOMAP. In addition, the collected data from physical environments also exist abnormal data (e.g., interference, incomplete and useless data), which can be filtered by utilizing PCA or ISOMAP.

On the other hand, data analytics is responsible for intelligently analyzing the collected data to discover useful information and form valuable knowledge. In 6G networks, massive data are collected from physical environment, cyber world, and social network which contain valuable information and meaningful features. Data analytics has brought us an excellent opportunity to understand the essential characteristics of wireless networks, and achieve more clear and in-depth knowledge of the behavior of 6G networks, finally valuable patterns or rules can be discovered as knowledge to provide suitable solutions for resource management, protocol adaptation, architecture slicing, cloud computing, signal processing, and so on. For instance, based on the discovered knowledge, ISAGUN is able to efficiently understand the mobility patterns of UAVs in the sky, establish the channel path loss model of satellite–ground link, and predict the device behavior in ground networks.

*C. Intelligent Control Layer*

Briefly, intelligent control layer mainly consists of learning, optimization, and decision-making, where this layer utilizes the appropriate knowledge from lower layers to enable massive agents (e.g., devices and BSs) to smartly learn, optimize and choose the most suitable actions (e.g., power control, spectrum access, routing management, and network association), with dual functions to support diverse services for social networks. Such function is realized by applying AI techniques in 6G networks, where each agent is equipped with an intelligent brain (learning model) to automatically learn to make decisions by itself.

Learning is the process of utilizing or modifying existing knowledge and experience to improve the behavior of each device or service center, so that 6G networks can intelligently realize optimal network slicing, end-to-end PHY design, edge computing, resource management, heterogeneous network design and so on according to different requirements of applications. Intelligence is the important characteristic of 6G networks, where the combination of AI and 6G networks can learn to achieve self-configuration, self-optimization, self-organization and self-healing, finally increasing the feasibility level. For instance, post-massive multiple-input multiple-output (PM-MIMO) will be employed in 6G networks to support hundreds or thousands of transmit/receive antennas with mmWave or THz transmissions [4]. How to achieve optimal energy-efficient beamforming variables and mitigate RF nonlinearities is challenging. An RNN-based solution has the ability to capture the nonlinearities of RF components, where RNN learns the nonlinearities of power amplifier arrays and optimizes minimal transmitted power levels at transmitters [11].

The task of optimization is to run a deterministic and rule-based network with parameter optimization based on global objectives (e.g., QoS, QoE, connectivity, and coverage). Traditional optimization algorithms (e.g., Lagrangian duality and gradient methods) have heavy mathematical models which may not be suitable for 6G networks, since 6G networks will be significantly dynamic and complex. In AI-enabled intelligent 6G networks, network parameters and architectures can be optimized through AI techniques instead of traditional tedious computation. AI techniques provide the best opportunity to train auto-learning models to realize network optimization for 6G wireless networks, allowing providers or operators to optimize the network parameters, resources or architectures to better adapt services for devices, finally make 6G networks become intelligent, agile, and able to adapt themselves. For instance, deep learning can enable SDN/NFV into an intelligent network architecture with fast learning, quick adaptation and self-healing, which is capable of quickly optimizing network parameters and architectures to achieve intelligent softwarization, cloudization, virtualization, and slicing.

Decision-making is an important cognitive task that enables massive agents to intelligently reason, plan, and choice the most suitable decisions to meet the high-quality service requirements. In decision-making, each agent that simultaneously attempts to obtain the new knowledge of other machines (called "exploration") and select the global actions with the highest benefit based on existing knowledge (called "exploitation"). The objective of decision-making is common in 6G networks, e.g., selecting the optimal precoding variables in mmWave or THz transceiver systems, choosing the suitable routing strategy for dynamic ISAGUN, and selecting the flexible spectrum management framework for massive multi-access scenario, all these decision-making issues can be effectively achieved by using AI techniques (e.g., MDP, game theory, RL, and optimization theory).

*D. Smart Application Layer*

The main responsibilities of this layer are delivering application specific services to the human beings according to their colorful requirements, as well as evaluating the provisioned services before feedbacking the evaluation results to the intelligence process. Intelligent programming and management can be achieved by the impetus of AI to support more various high-level smart applications, such as automated services, smart city, smart industry, smart transportation, smart grid and smart health, and handle global management relevant to all smart type applications. All the activities of smart devices, terminals and infrastructures in 6G networks are also managed by the smart application layer through the AI techniques to realize network self-organization ability.

Another objective of this layer is to evaluate the service performance, where lots of considerations and factors can be involved, such as QoS, QoE, quality of collected data, and quality of learned knowledge. At the same time, the cost dimension metric in terms of resource efficiency is also required to be taken into account, such as spectrum utilization efficiency, computational efficiency, energy efficiency and storage efficiency. All the above mentioned evaluation metrics can be utilized to improve intelligent resource management, automatic network slicing, and smart service provisioning.

## III. ARTIFICIAL INTELLIGENCE TECHNIQUES FOR 6G NETWORKS

In this section, we will elaborate how the AI techniques grant preliminary intelligence for 6G networks in terms of AI-empowered mobile edge computing, intelligent mobility and handover management, and smart spectrum management.

### A. AI-Empowered Mobile Edge Computing

Mobile edge computing (MEC) will be an important enabling technology for the emerging 6G networks, where MEC can provide computing, management and analytics facilities inside RAN or SDN in close proximity to various devices. In MEC networks, the decision making optimization, knowledge discovery and pattern learning are sophisticated due to the multi-dimensional, randomly uncertain, and dynamic characteristics. Hence, traditional algorithms (e.g., Lagrangian duality) may face the limitation in such complex networks. AI techniques can extract valuable information from collected data, learning and supporting different functions for optimization, prediction, and decision in MEC [12]. Fig. 3 shows the framework of AI-empowered mobile edge computing, which consists of central cloud computing and edge computing.

In edge computing servers, due to the limited capability, lightweight AI algorithms can be utilized to provide smart applications for edge scenarios (e.g., transportation and agriculture), as shown in Fig. 3(b). For example, RL-based edge computing resource management is a model-free scheme which does not need historical knowledge and it can be used to learn the environment dynamics and make suitable control decisions in real-time [13]. In the RL framework, at each step, after obtaining the state (e.g., device mobility, requirement dynamics, and resource condition) by interacting with environments, the possible resource management solutions (e.g., energy management, resource allocation, and task scheduling) are contained into the set of possible actions. Each RL agent (e.g., device or service center) selects the best action from a set of possible actions or chooses one action randomly to maximize its reward, where the reward can be determined by data rate, latency, reliability, and so on.

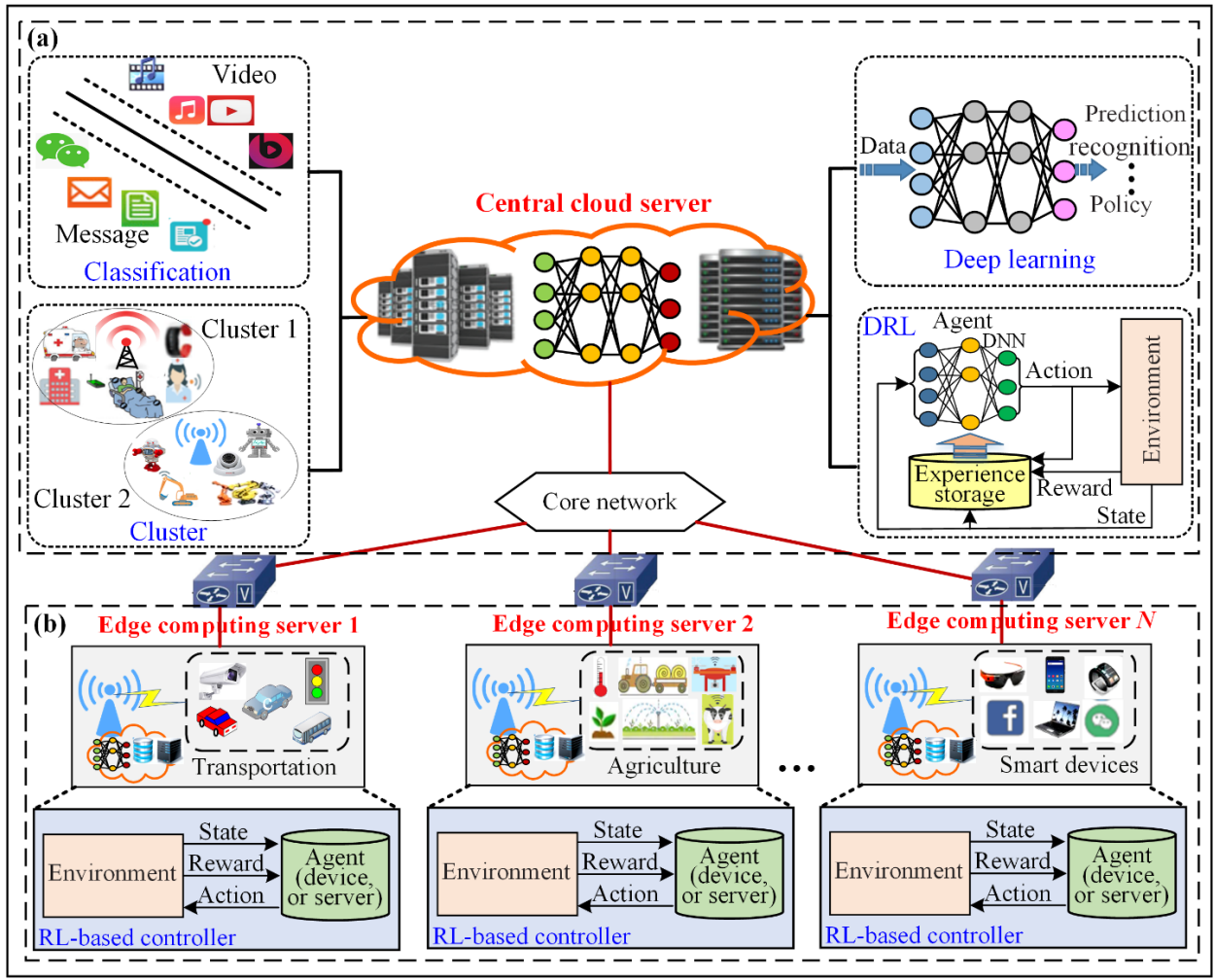

Figure 3. The framework of AI-empowered MEC.

In the central cloud server, since it has powerful computation capability, complex centralized large-scale AI algorithms can be employed to provide various learning functions [12], as shown in Fig. 3(a). For instance, as service applications in MEC networks are diverse and dynamic, AI-based classification can be used to efficiently customize traffic flow decision for various service features. In addition, MEC server association can be obtained by AI-based cluster instead of individual decision, which will be more effective to reduce a number of participants greatly. The central cloud server may receive massive data from edge computing servers and the data need to be trained to automatically extract features and discover knowledge. In this case, deep learning can be adopted to train computational models to achieve service recognition, traffic and behavior prediction, security detection, and so on. Moreover, in complex and dynamic MEC networks, the mapping between resource management decisions and the effect on the physical environments is not easy to be analytically defined. DRL can be adopted to search the optimal resource management policy under high-dimensional observation spaces. Experience replay is also adopted in DRL to utilize the historical knowledge to improve learning efficiency and accuracy, allowing the MEC to support high-quality services for edge devices.

*B. Intelligent Mobility and Handover Management*

Mobility and handover management are probably two most challenging issues of 6G networks, since 6G networks will be high dynamic, multi-layer, and large-dimensional, leading to frequent handovers. AI techniques can be adopted to intelligently achieve mobility prediction and optimal handover solutions to guarantee communication connectivity [8], [10].

For example, UAVs communications will be integrated into 6G networks, and the high-speed mobility of UAVs may lead to frequent handovers. In addition, the diverse service requirements in terms of high data rate, high reliability and low latency increase the difficulty in processing efficient handover. At the same time, the high mobility of devices and UAVs results in uncertainties of their locations. One of AI technologies, namely, DRL (DRL combines DL with RL to learn itself from experience), is capable of solving complex decision-making tasks, which learns to optimize the handover strategies in real-time by exhibiting dynamic temporal mobility behaviors of devices or UAVs in an online manner, while minimizing the transmission latency and guaranteeing reliable wireless connectivity [10]. Fig. 4 shows the context of intelligent mobility and handover management based on DRL for UAV-enabled networks, where each UAV can be regarded as a learning agent to learn management policy by interacting with its environments. Each agent senses the environment states (e.g., link quality, current location, and velocity) and discovers the most suitable actions (e.g., mobility and handover parameters) to obtain the greatest reward, where the reward can be determined by communication connectivity, latency, capacity and so on. In the DRL framework, UAVs can learn how to move and handover automatically and robustly, how to reduce the latency and the handover failure probability, finally provide better services for ground devices.

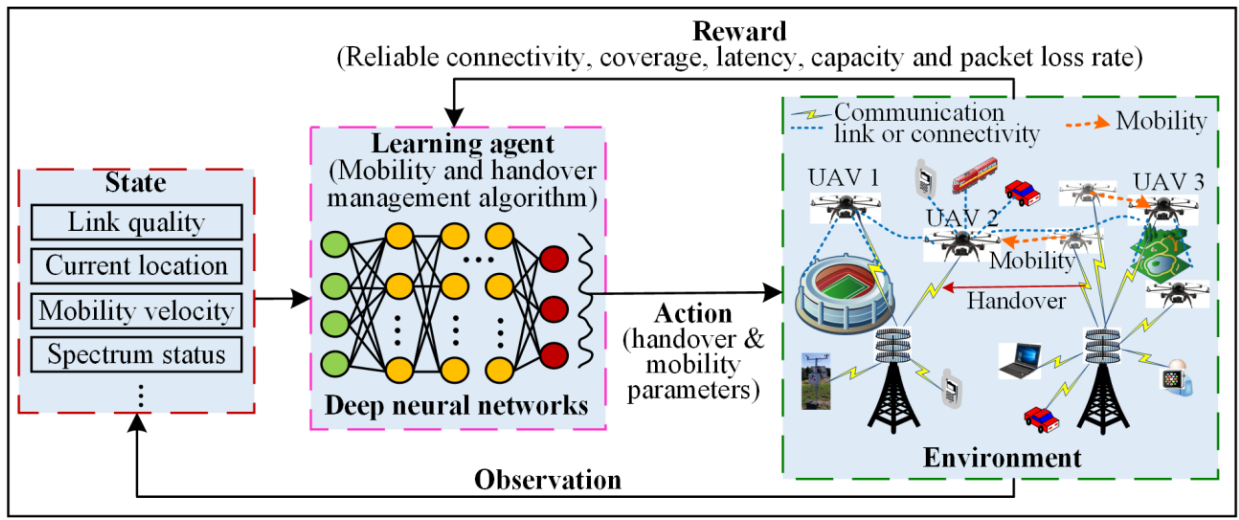

Figure 4. DRL in the context of mobility and handover management. Note: mobility and handover parameters include BS association, spectrum access, and trajectory [10].

6G networks need to meet high-speed mobility and delay-sensitive requirements of vehicles in large-scale vehicular networks, so efficient mobility management is a key evaluation to satisfy reliability, continuity and low latency requirements of vehicular communications. Deep learning (such as, RNN and ANN) based predictive mobility management and fuzzy Q-learning based handover parameter optimization can learn the mobility patterns of high-speed vehicular users, which can effectively void frequent handovers, handover failures or connectivity failures [14]. In addition, LSTM is also a powerful AI tool for solving handover problems, as it exploits both the previous and future mobility contexts of vehicles for learning a sequence of future time-dependent movement states, finally predict vehicles' trajectories to optimize handover parameters to avoid frequent handovers.

## C. Intelligent Spectrum Management

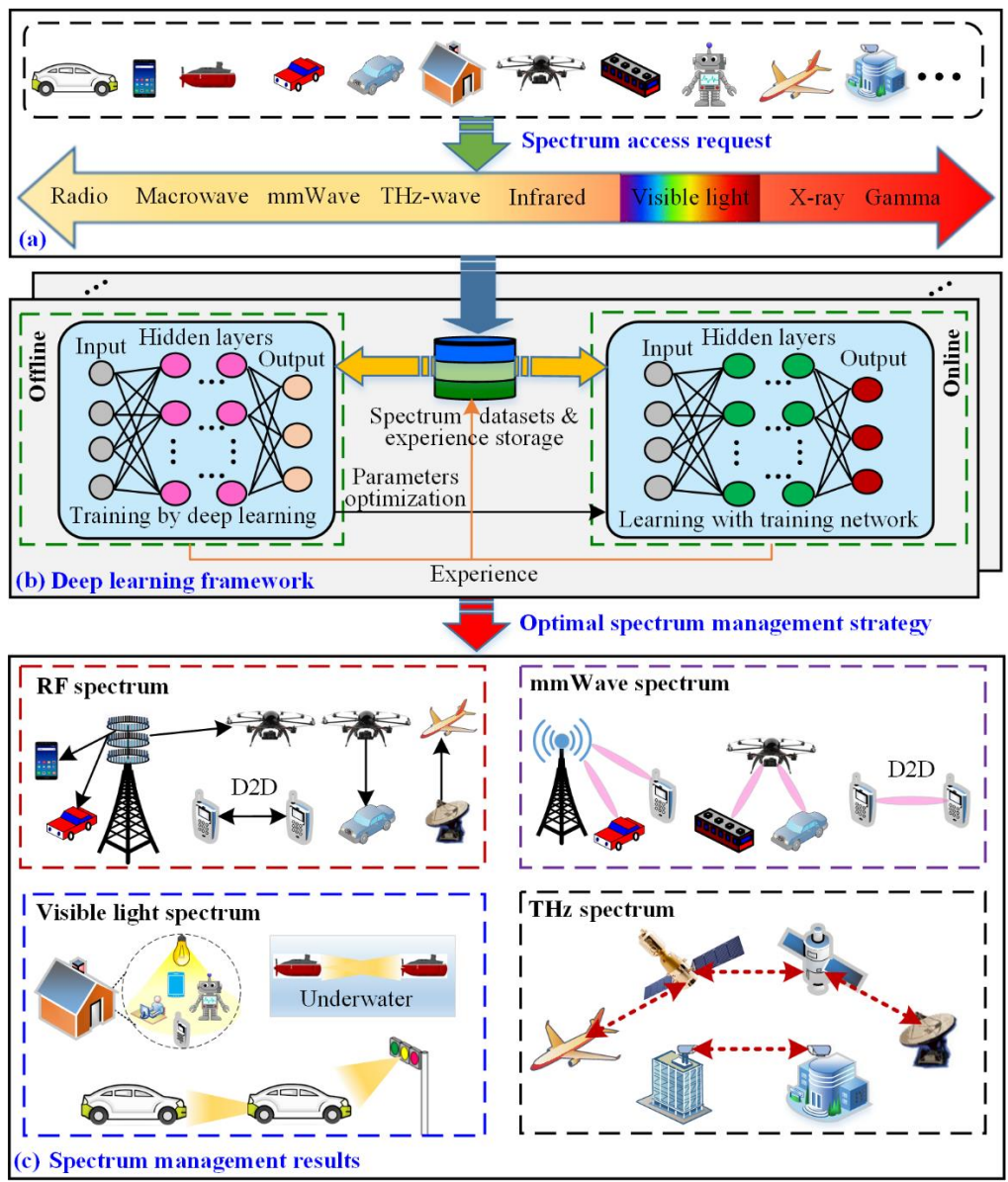

Figure 5. Deep learning-based flexible spectrum management framework.

6G networks utilize different spectrum bands (e.g., low radio frequency, mmWave, THz, and visible light spectrum) to support high data rates, as shown in Fig. 5 (a). When a massive number of devices are involved in 6G networks to require spectrum assignment, AI-enabled spectrum management is capable of intelligently supporting massive connectivities and diverse services, as shown in Fig. 5. The AI-enabled learning framework constrains three layer manners, namely input layer, hidden or training layer and output layer, as shown in Fig. 5 (b), where the framework first modules input massive spectrum datasets in different radio contexts [15]. It then trains the hidden layers by comprehensively analyzing the characteristics of current or previous spectrum utilization information and discovers meaningful spectrum usage characteristics. Finally, the most suitable spectrum management strategies are provided in the output layer in real-time to support massive connectivities for devices.

The learning framework also constrains an offline training model that the trained models, past experience and developed rules can be stored to smoothly realize online spectrum management decisions. With the help of AI techniques, different spectrum bands can be smartly utilized for different types of traffic transmissions for satisfying colorful service requirements, such as visible light and THz bands can be exploited for high-capacity transmissions with large bandwidth, while low-frequency bands can be assigned to broadcast short messages for satellite–ground transmissions.

## IV. Future Research Directions

AI can optimize the network performance based on its powerful learning ability and strong reasoning ability. In this section, we outline some future research topics and potential solutions for AI-enabled intelligent 6G networks.

**Computation Efficiency and Accuracy:** In 6G networks, the massive amounts of collected data and complex network architectures pose challenges for AI-enabled learning and training process. In addition, the limited computing resources may be insufficient to process massive high-dimensional data to meet the training accuracy rate. Furthermore, deep learning generally has high computational complexity, which is prohibitively expensive. Hence, how to design efficient AI learning schemes to improve both the computation efficiency and accuracy pose a significant research challenge. Recently, residual networks, graphics processing, feature matching, and offline training are being considered as promising methods with high performance computing facilities for encouraging convergence speed, reducing complex computations, and improve training accuracy.

**Robustness, Scalability, and Flexibility of Learning Frameworks:** As we mentioned above, 6G networks (e.g., vehicular networks and UAV-enabled networks) exhibit high dynamics in some facets, such as BSs associations, wireless channels, network topologies, and mobility dynamics. In particular, devices or terminals who join or leave networks may have different QoS and QoE requirements. All these mentioned uncertainties in dynamic networks call for ongoing updates of parameters of AI learning algorithms. Great robustness, scalability and flexibility of learning frameworks are crucial aspects for supporting the potential unbounded number of interacting entities and providing high-quality services in real-world dynamic networks. Thus, how to design robust, scalable, and flexible learning frameworks for 6G networks is still an open issue.

**Hardware Development:** The corresponding hardware development is very challenging when designing 6G networks. On the one hand, hardware components have high energy consumption and expensive cost when they work at mmWave and THz bands. On the other hand, some devices/terminals have limited storage and computing energy. Despite benefits of AI learning algorithms in terms of learning and recognition ability, they usually require high computational complexity, power consumption, and sufficient computing resource. Hence, collaboration among hardware components and AI learning algorithms need to be advocated, which needs significant research efforts on it. It is worth noticing that graphics processing unit (GPU) is suitable for learning computation as it can effectively handle matrix computations, and transfer learning-enabled intelligent communications that can adapt to different hardware constraints by transferring one learning framework to other hardware components.

**Energy Management:** In 6G networks, some devices located on the ground can be charged by power stations, but undersea, air, and space infrastructures as well as some sensors cannot be connected to power stations. Moreover, 6G networks also need to flexibly connect the massive number of smart low-power devices. In this case, advanced energy management schemes become very critical in 6G networks. AI techniques have ability to help these infrastructures and devices to optimize energy management strategy by intelligently managing their energy consumption or harvesting energy (e.g., energy harvesting and wireless power transfer), hence increase the usage time. Therefore, energy management is an important but challenging topic for 6G networks.

## V. CONCLUSION

In this article, we have proposed an AI-enabled intelligent architecture for 6G networks by utilizing AI techniques, with the purpose of smartly supporting diverse services, optimizing network performance and guaranteeing seamless connectivity. We then presented some AI-enabled applications for addressing different aspects of 6G networks deployment and management, including AI-empowered mobile edge computing, intelligent mobility and handover management, and smart spectrum management. Finally, we highlighted several promising research directions and potential solutions for 6G networks.

## REFERENCES

[1] J. Andrews *et al.*, "What Will 5G Be?" *IEEE J. Sel. Areas Commun.*, vol. 32, Jun. 2014, pp. 1065–82.

[2] F. Boccardi, R. W. Heath, A. Lozano, T. L. Marzetta, and P. Popovski, "Five Disruptive Technology Directions for 5G," *IEEE Commun. Mag.*, vol. 52, no. 2, Feb. 2014, pp. 74-80.

[3] K. David and H. Berndt, "6G Vision and Requirement," *IEEE Vehic. Teh. Mag.*, vol. 13, no. 3, Sept. 2018, pp. 72–80.

[4] P. Yang, Y. Xiao, M. Xiao, and S. Li, "6G Wireless Communications: Vision and Potential Techniques," *IEEE Network*, vol. 33, no. 4, Jul. 2019, pp. 70-75.

[5] K. B. Letaief, W. Chen, Y. Shi, J. Zhang, and Y. A. Zhang, "The Roadmap to 6G: AI Empowered Wireless Networks," *IEEE Commun. Mag.*, vol. 57, no. 8, Aug. 2019, pp. 84-90.

[6] Z. Zhang *et al.*, "6G Wireless Networks: Vision, Requirements, Architecture, and Key Technologies," *IEEE Vehic. Teh. Mag.*, vol. 14, no. 3, Sept. 2019, pp. 28-41.

[7] M. G. Kibria, K. Nguyen, G. P. Villardi, O. Zhao, K. Ishizu, and F. Kojima, "Big Data Analytics, Machine Learning, and Artificial Intelligence in Next-Generation Wireless Networks," *IEEE Access*, vol. 6, May 2018, pp. 32328-32338.

[8] N. Kato *et al.*, "Optimizing Space-Air-Ground Integrated Networks by Artificial Intelligence," *IEEE Wireless Commun.*, vol. 26, no. 4, Aug. 2019, pp. 140-147.

[9] S. J. Russell and P. Norvig, Artificial Intelligence - A Modern Approach. *Pearson Education*, 2010.

[10] C. H. Liu, Z. Chen, J. Tang, J. Xu, and C. Piao, "Energy-Efficient UAV Control for Effective and Fair Communication Coverage: A Deep Reinforcement Learning Approach," *IEEE J. Sel. Areas Commun.*, vol. 36, no. 9, Sep. 2018, pp. 2059-2070.

[11] M. Yao, M. Sohul, V. Marojevic, and J. H. Reed, "Artificial Intelligence Defined 5G Radio Access Networks," *IEEE Commun. Mag.*, vol. 57, no. 3, Mar. 2019, pp. 14-20.

[12] B. Cao, L. Zhang, Y. Li, D. Feng and W. Cao, "Intelligent Offloading in Multi-Access Edge Computing: A State-of-the-Art Review and Framework," *IEEE Commun. Mag.*, vol. 57, no. 3, Mar. 2019, pp. 56-62.

[13] A. Nassar and Y. Yilmaz, "Reinforcement Learning for Adaptive Resource Allocation in Fog RAN for IoT With Heterogeneous Latency Requirements," *IEEE Access*, vol. 7, 2019, pp. 128014-128025.

[14] H. Yang, X. Xie, and M. Kadoch, "Intelligent Resource Management Based on Reinforcement Learning for Ultra-Reliable and Low-Latency IoV Communication Networks," *IEEE Trans. Vehic. Teh.*, vol. 68, no. 5, May 2019, pp. 4157-4169.

[15] U. Challita, L. Dong, and W. Saad, "Proactive Resource Management for LTE in Unlicensed Spectrum: A Deep Learning Perspective," *IEEE Trans. Wireless Commun.*, vol. 17, no. 7, Jul. 2018, pp. 4674-4689.